# TRANSVERSE INSTABILITIES IN THE FERMILAB RECYCLER

L. R. Prost, A. Burov, A. Shemyakin, C. M. Bhat, J. Crisp, N. Eddy,
FNAL, Batavia, IL 60510, U.S.A.

*Abstract*

Transverse instabilities of the antiproton beam have been observed in the Recycler ring soon after its commissioning. After installation of transverse dampers, the threshold for the instability limit increased significantly but the instability is still found to limit the brightness of the antiprotons extracted from the Recycler for Tevatron shots.

In this paper, we describe observations of the instabilities during the extraction process as well as during dedicated studies. The measured instability threshold phase density agrees with the prediction of the rigid beam model within a factor of 2. Also, we conclude that the instability threshold can be significantly lowered for a bunch contained in a narrow and shallow potential well due to effective exclusion of the longitudinal tails from Landau damping.

## 1. INTRODUCTION

Fermilab's Recycler was designed to provide an additional storage ring for the accumulation of 8 GeV antiprotons [1] and is now a critical component of the accelerator complex. In the effort to provide higher integrated luminosity for the experiments, the number of antiprotons stored in the Recycler continually increased and reached up to $500 \times 10^{10}$ particles. At this level ($\sim 400 \times 10^{10}$ and higher), the antiproton beam is subject to a transverse instability [2,3,4] during the RF manipulation [5] necessary for extraction to the Tevatron, while cooling is also the strongest. In other words, the transverse instability sets the maximum brightness of the antiproton beam that can be delivered to the downstream machines and in turn the overall efficiency for the number of antiprotons available for collision.

This paper is organized in the following manner: we (1) give an overview of the instability theoretical model, and introduce the 'phase density' parameter $D$ [6] used during normal operation to determine the onset of an instability; (2) summarize the use and upgrades of the dampers installed to alleviate instabilities; (3) give an account of the instabilities that occurred during normal operation and show a typical example; (4) present dedicated studies carried out to experimentally determine the instability phase density threshold for various RF configurations; (6) discuss our results and propose methods that could improve further the beam stability during normal operation; (7) conclude.

## 2. THEORETICAL FRAMEWORK

### 2.1 Phase density instability threshold

The instability the Recycler experiences is related to the antiproton beam own space charge, which separates coherent and incoherent betatron frequencies. In turns, it drastically reduces Landau damping, which is a transfer of energy from the coherent motion to the incoherent oscillations of the resonant particles (*i.e.* particles whose individual betatron frequencies are identical to the coherent frequency). As a consequence, even a tiny impedance would drive an instability.

In the context of the rigid beam model, for a coasting beam with Gaussian distributions (both longitudinally and transversely) and assuming that the main reason for the frequency spread (for the resonant particles) is the chromaticity, the stability threshold follows [4]:

$$\frac{\Delta \nu_{sc}}{\sigma_{\nu p}} = 1.7 \ln\left(\frac{\Delta \nu_{sc}}{\operatorname{Im}\Delta \nu_c}\right), \tag{1}$$

where $\sigma_{\nu p} = |\xi - n\eta| \sigma_p \equiv \xi_n \sigma_p$ is the effective chromatic rms tune spread for mode $n$, $\eta$ is the slippage factor, $\sigma_p = \Delta p_{rms} / p$ is the relative momentum rms spread, $\Delta \nu_{sc}$ is the maximal space charge tune shift

$$\Delta \nu_{sc} = \frac{r_p \lambda C}{4\pi \beta \gamma^2 \varepsilon_{T\,nrms}} \tag{2}$$

with $\lambda$ as the linear density, $C$ the orbit circumference and $r_p$ the proton classical radius, $\varepsilon_{T\,nrms}$ is the rms normalized transverse emittance and $\Delta \nu_c$ is the wake-driven coherent tune shift, see *e. g.* [7]:

$$\Delta \nu_c = -i \frac{\lambda r_p}{\gamma} \frac{\langle \beta_\perp Z_\perp \rangle}{Z_0} \tag{3}$$

with $\langle \beta_\perp Z_\perp \rangle$ as the ring-averaged product of the transverse impedance and beta-function, $Z_0 = 4\pi / c = 377\,\text{Ohm}$. For a flat vacuum chamber with a half-gap $b$, the transverse impedance can be written as

$$Z_\perp(\omega) = Z_0 F_Y \frac{C\delta(\omega)(1 - i\,\text{sign}\,\omega)}{2\pi b^3};$$

$$F_Y = \begin{cases} \pi^2 / 12, & \text{vertical;} \\ \pi^2 / 24, & \text{horizontal.} \end{cases} \tag{4}$$

Here $\delta(\omega) = c / \sqrt{2\pi \sigma_c \omega}$ is the skin depth with $\sigma_c$ as the chamber conductivity (assumed to be stainless steel, $\sigma_c = 1.3 \cdot 10^{16}\,\text{s}^{-1}$ hereafter); $F_Y$ is the geometry Yokoya factor (for a round chamber $F_Y = 1$.)

It should be noted that, the model presented here [4] uses a more realistic beam particle distribution for space-charge tune shift calculations than a similar model from Ref. [3] and gives ≈ 40% higher instability threshold.

For operational purposes, an effective phase density [6] is defined as:

$$D_{95} = \frac{N_{\bar{p}}}{4\varepsilon_{L\,rms} \cdot 6\varepsilon_{T\,nrms}} \quad (5)$$

where $N_{\bar{p}}$ is the number of antiprotons in units of $10^{10}$, $\varepsilon_{L\,rms}$ is the rms longitudinal emittance in eV s and $\varepsilon_{T\,nrms}$ is the rms. normalized transverse emittance in μm. The numerical factors are chosen to give 95% emittances for the Gaussian distributions. Then, following Eqs. (1-5), one can rewrite the instability threshold in terms of this effective phase density:

$$D_{th,95} = 60F\frac{\gamma_0^2 \xi_n}{T_0[s]E_0[eV]}, \quad F \equiv \ln\left(\frac{\Delta\nu_{sc}}{\mathrm{Im}\,\Delta\nu_c}\right); \quad (6)$$

units for the revolution time $T_0$ and the beam energy $E_0 = \gamma_0 m_p c^2$ are shown in the brackets. Note that this threshold depends on the mode frequency $f = (n - \{\nu_b\})f_0 = (n - \{\nu_b\})c/C$, where $\{\nu_b\}$ is a fractional part of the betatron tune. In case there is no damper, a mode with the lowest threshold density determines actual threshold; for the resistive wall impedance it is a lowest unstable mode, $f = (1 - \{\nu_b\})f_0$. In case there is a damper, the threshold is determined by its bandwidth. Figure 1 illustrates how the coasting beam threshold density depends on the damper's upper frequency, or the bandwidth for the Recycler ($\eta = -0.0085$, $\xi = -6$) assuming a typical 95% emittance for extraction of 3 π mm mrad normalized.

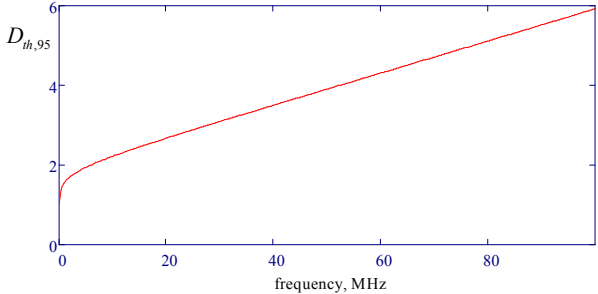

Figure 1: Threshold effective density for a coasting beam versus the damper's upper frequency $f$.

## 2.2 Predicted instability thresholds for the Recycler

For the Recycler, a calculation of $D_{th,\,95}$ gives ~1.0 at the lowest sideband ($n = 1$) and the beam parameters used for the computation of the values in Table 1. Hence, without any external damping, and reasonable bunching ratio, it limits the number of antiprotons that can be stored to ~220×$10^{10}$. Note that in this case, shortening the bunch increases the beam stability although very weakly [2, 3]. As a remedy, a transverse digital damper system was installed in 2005 [8].

Originally, the dampers' bandwidth was limited to ~30 MHz (i.e. $n \sim 330$), increasing the stability region of Eq. (6) to $D_{th,95}$ =3.1. However, to accommodate the need for higher antiproton brightness, the original system was upgraded, and the operational bandwidth increased from 30 MHz to 70 MHz [9, 10], bringing the phase density instability threshold to ~4.7. Currently, during normal operation and storing conditions, we find that the beam remains stable up to $D_{95} \sim$ 3.5-7.0, in line with expectations. While increasing the dampers bandwidth by a factor of 2 was entirely done through improving the electronics, going further would require hardware modifications in the vacuum chamber (kickers and/or pickups). There are no such plans for the remaining of the Tevatron running period.

Table 1: Summary table of the thresholds calculated with Eq. (6). $\xi$ = -6; $6\varepsilon_{T,\,rms}$ = 3 π mm mrad.

|  | $Z_\perp(\omega_n)/C$ [MΩ m$^{-1}$] | $\ln\left(\frac{\Delta\nu_{sc}}{\mathrm{Im}\Delta\nu_c}\right)$ | $D_{th,95}$ |
|---|---|---|---|
| No dampers ($n$ = 1) | 28 | 3.1 | 1.0 |
| 30 MHz dampers ($n$ ~330) | 1.2 | 6.3 | 3.1 |
| 70 MHz dampers ($n$ ~ 780) | 0.8 | 6.7 | 4.7 |

$\omega_n$ ≡ coherent frequency for mode $n$.

## 2.3 Importance of the beam distribution in the determination of the instability threshold

In the preceding section, all calculations have been carried out using the example of a Gaussian distribution in order to simplify the results. However, it should be noted that since the antiproton beam in the Recycler is not exactly known (or Gaussian), the instability threshold values presented should not be entirely relied upon. The role that the beam distribution plays in the determination of the instability threshold may be seen through the calculation of the Landau damping rate. A general expression is given by Eq. (6) of Ref. [4]. Assuming once again that the tune spread is due to chromaticity only and that the space charge tune shift is a constant (i.e. does not depend on the transverse actions or longitudinal position), the Landau damping rate, Λ, can be expressed as:

$$\Lambda \cong \frac{\pi\langle\Delta\nu_{sc}\rangle^2}{|\xi_n|}\hat{f}\left(\frac{\langle\Delta\nu_{sc}\rangle}{|\xi_n|}\right); \quad \langle\Delta\nu_{sc}\rangle \cong 0.5\Delta\nu_{sc} \quad (7)$$

where

$$\hat{f}\left(\frac{\Delta p}{p}\right) = \iint f\left(J_x, J_y, \frac{\Delta p}{p}\right)dJ_x dJ_y \quad (8)$$

In Eq. (8), the distribution function $f$ is arbitrary and, $J_x$ and $J_y$ are the transverse actions. Since Landau damping is the only stabilizing mechanism (without dampers), when Λ goes to zero, the beam goes unstable. Therefore, in theory, since the Landau damping rate is proportional to the beam distribution function integrated over its transverse actions, it is necessary to know the details of the distribution $f$ to determine if the beam will go unstable or not. For instance, calculations of $D_{th,95}$ for a Gaussian and a step-like distribution find that it is ~2 times higher for the Gaussian distribution than for the step-like distribution [4]. Unfortunately, in operation, there is no measurement that can resolve quantitatively the amount of tail particles so that we could exactly predict (thus avoid) the instability to occur.

## 2.4 Bunching effects

### 2.4.1 Barrier bucket with infinite walls

Typically, the antiproton beam in the Recycler is contained between two rectangular RF barriers. For a bunch with negligible synchrotron tune, the tail-to-head interaction takes place due to a long-range wake field. This leads to a dependence of the coherent tune shift $\Delta\nu_c$ on the bunching factor $B = T_0/\tau$, where $\tau$ is the bunch length. The Recycler's wake field is believed to be dominated by the resistive wall contribution; thus the coherent tune shift slowly grows when the bunch length decreases; for a single bunch in the ring $\text{Im}(\Delta\nu_c) \propto B^{1/3}$ [3], close to a two-particle model where $\Delta\nu_c \propto B^{1/4}$ [2]. In turn Eq. (6) contains a logarithmic dependence on the bunching factor mostly due to $\Delta\nu_{sc} \propto B$. Figure 2 shows how the threshold density depends on the bunching factor, with a 70 MHz damper system for the beam in a barrier bucket of infinite height.

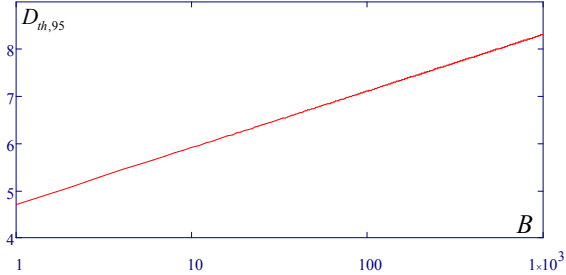

Fig. 2: Threshold effective density versus bunching factor, with 70 MHz damper on, and the same emittance as for Figure 1. The potential well is deep so as to contain all the resonant particles.

### 2.4.2 Effect of the finite depth of the potential well

According to Eq. 7, the resonant particles belong to a surface in the $(\Delta p, J_x, J_y)$ space which is determined by the condition of equality of these particles' tunes to the coherent tune:

$$\xi_n \frac{\Delta p}{p} + \Delta Q_{sc}(J_x, J_y) = 0 \qquad (9)$$

The exact shape of this surface depends on the transverse particle distribution. As an example, a solution to Eq. (9) is shown in Figure 3 for a Gaussian round beam where $\Delta Q_{sc}$ is calculated with Eq. (8) in Ref. [4]. Along the $J_x$ and $J_y$ axes, the surface extent is limited by the accelerator aperture. Consideration of the momentum offset is more complicated.

The maximum momentum offset $\Delta p_{res\_max}$ of the resonant particles is at the beam center

$$\Delta p_{res\_max} = \left| \frac{\Delta Q_{sc}(0,0)}{\xi_n} p \right| \qquad (10)$$

Note that for the Recycler case (negative chromaticity), only particles with a negative momentum offset participate in Landau damping.

If the barrier height were infinite, the maximum value of $\Delta p$ at the resonant surface would be determined by either the Recycler momentum aperture or by the value of $\Delta p_{res\_max}$, calculated with Eq. (10).

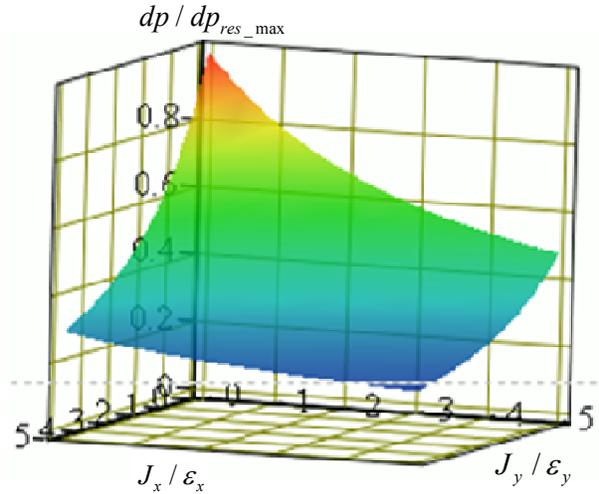

Figure 3: Surface of the resonant particles for oscillations in X. Horizontal axes correspond to transverse actions normalized by rms emittances. The vertical axis represents the momentum offset normalized by its resonant value at the beam center $\Delta p_{res\_max}$.

Finite RF barriers may lead to an effective loss of the resonant particles. Indeed, particles with energies greater than the bucket height escape and drift around the ring (here called DC particles). Figure 4 schematically shows a DC particle (red), and a captured particle (blue), which oscillates within the potential well. The DC particles accelerate and decelerate as they cross over the RF barriers. If one considers a particle that barely escapes the potential well, its "velocity" along the horizontal axis of Figure 4 is low. Correspondingly, the density of these particles inside the bucket is significantly decreased. On the other hand, outside of the bucket the resonant particles are coupled with the bunch by weak wake fields, while inside the bucket the coupling is provided by strong space charge fields. As a result, contribution of the DC particles into Landau damping is suppressed. Hence, in the simplest model, the barrier height sets an effective maximum for the momentum offset of the resonant particles. On Figure 3, this condition would be represented by a plane parallel to XY.

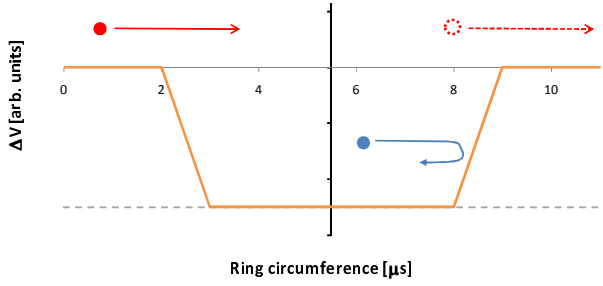

Figure 4: Simplified schematic of the potential well for a 'cold bucket' RF structure. The blue particle represents a captured particle, which oscillates within the potential well. The red particle represents a so-called DC particle, which drifts around the ring.

To further illustrate the meaning of the resonant particles surface of Figure 3, let us consider a particles distribution $f$ of the form

$$f(J_x, J_y, \Delta p) \propto e^{-\left(\frac{J_x}{2\varepsilon_x} + \frac{J_y}{2\varepsilon_y} + \frac{\Delta p^2}{2\sigma_p^2}\right)} \quad (11)$$

where, $\varepsilon_x$ and $\varepsilon_y$ are the transverse emittances, and $\sigma_p$ the rms momentum spread of the distribution. For realistic parameters, Figure 5 shows the intersection of the resonant particles surface with the ($J_x, \Delta p$) plane at $J_y = 0$ (red trace); the barrier height is then represented by a straight line at a fixed $\Delta p$ (blue trace); and the 3 brown dotted lines are the intersection of the distribution $f$ of Eq. (11) with the same plane.

In this simplified representation, the particles providing Landau damping are on the red line. It is then readily apparent that the contributing particles to Landau damping have either large momentum spread or large transverse actions, thus justifying the importance of the distribution tails in the determination of the stability limit. Note that the particles with low transverse actions are excluded from Landau damping because their momentum offset (while inside the bucket) is above the bucket height (represented by the blue line on the plot). Hence, this model predicts that the beam stability decreases if the barrier height is less than the value determined by Eq. (10). More quantitative predictions are difficult because, as shown, it strongly depends on the tail distribution, which is not known experimentally.

2.4.3 Other RF configurations

A third factor, which would alter the coasting beam model, is the possibility for the potential well profile to depart from the one resulting from a barrier RF configuration. Before extraction, the beam is kept inside cosine-like potential wells; hence the barrier-bucket theory does not apply. Similar to head-tail modes with strong space charge, where smooth walls of the potential well are better for Landau damping [8], the beam stability threshold for this case can be expected to increase as well.

However, it should be mentioned that the presence of multiple bunches around the Recycler, also affects the way an instability develops. Indeed, other bunches play the role of 'relay stations' for the tail-head signal, thus increasing the coherent growth rate, therefore logarithmically decreasing the instability threshold.

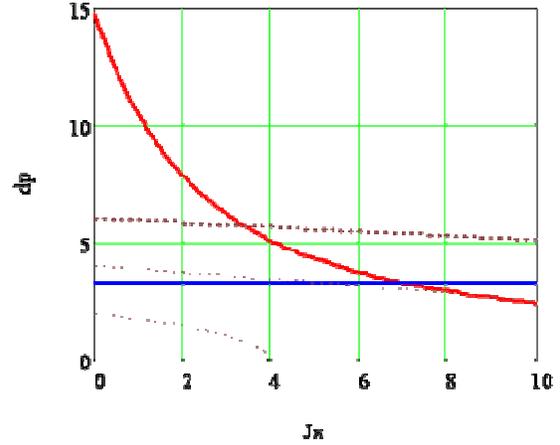

Figure 5: The intersection of the surface shown in Figure 3 with the plane ($J_x, \Delta p$) at $J_y = 0$ is represented by the red line. The transverse action is normalized by the emittance. In contrast to Figure 3, the momentum offset on the vertical axis is normalized by its rms. value $\sigma_p$. The horizontal blue line represents the maximum barrier height also normalized by $\sigma_p$. $\Delta p_{res\_max} = 38$ MeV/c is calculated for the parameters of Case (this paper nomenclature) B in Table 5 with Eq. (10) and assuming a Gaussian transverse distribution. The dotted brown lines show contours of the surfaces of equal phase density intersection the resonant particles surface at $\Delta p/\sigma_p = 2$, 4, and 6.

2.4.4 Summary

During normal operation in the Recycler, the bunch length and RF structure, with its imperfections (*i.e.* deviations from the nearly ideal waveforms generated by the low level RF system), vary. The models presented above give qualitative answers to the effect these manipulations may have on the beam stability (except the infinite wall model for which $D_{th,95}$ was explicitly computed as a function of $B$).

For instance, when the bunch is being compressed (shortened, *i.e. B* increases), the infinite wall model predicts that the beam becomes more stable. On the other hand, when the depth of the potential well is finite, compressing the bunch decreases the instability threshold, since resonant particles spill outside the potential well and become DC. At the same time, the details of the potential well, deviations from ideality for the barrier buckets configuration, and the various RF structures (during injections to and extractions from the Recycler) also play significant roles in determining how the instability threshold evolves. In short, as it was stated previously, predictions are difficult.

Nevertheless, experimental results (during operation or studies) confirm the finite depth potential well model description: shortening the bunch does decrease the beam stability.

It is important to mention here that within the framework of the theoretical model developed and used in this paper, the potential well distortion, or beam loading for bunches in rectangular barrier buckets, are not taken into account, since any distortion in the bunch profile is corrected by means of a linearization circuit and a FPGA based adaptive beam loading compensation system [11].

## 2.5 Consequence from the Recycler having an elliptical vacuum chamber and uncertainties on chromaticities

The theory presented up to now does not differentiate between the two transverse degrees of freedom *i.e.* the value of $D_{th,\,95}$ is applicable for both the horizontal and vertical directions. However, all instabilities experienced with properly working dampers were observed in the vertical plane.

The most probable explanation is based on the fact that the vertical resistive wall impedance is a factor of 2 higher than the horizontal. The main reason for this asymmetry is because the Recycler beam pipe is mostly elliptical around the ring. Thus, for identical chromaticities and damper bandwidths, the horizontal instability cannot be seen, since the vertical threshold is slightly lower due to the logarithmic factor $F$ in Eq. (6). On the other hand, this slight logarithmic difference can be outweighed by a small difference in the effective chromaticities $\xi_{nx}$ and $\xi_{ny}$ (defined in Section 2.1) if the absolute value of the vertical chromaticity sufficiently exceeds that of the horizontal. When the normal chromaticities $\xi_{x,y}$ are small, and the effective chromaticities are dominated by the longitudinal factor $n\eta$, the polarization of the instability depends on an interplay of these two weak factors, and may spontaneously change due to a slight uncontrolled variation in the chromaticities.

## 3. EARLY HISTORY

### 3.1 Before dampers implementation

Sources of instabilities in the Recycler have been theoretically studied during its design (for instance in [12]) but were deemed a marginal issue for the maximum number of antiprotons that were supposed to be stored at any time ($< 250 \times 10^{10}$). While installing octupole magnets was considered to improve Landau damping, they were eventually not included in the final design.

Since its first use with antiprotons, which allows using the stochastic cooling system, instabilities have been observed. An ion-capture driven instability was soon identified and was eliminated with clearing electrodes and the fact that the stored beam was bunched.

The first dedicated studies with antiprotons were carried out in 2004 (*e.g.*: June 9, June 21 and July 8). At that time, instabilities were typically induced by reducing the Recycler chromaticity (with stochastic cooling on) and signals from BPMs were used to characterize it. These measurements and others performed in 2005 were the basis for the specifications of the digital dampers which were installed during the summer 2005 (July) and fully commissioned in the fall (October) [8].

### 3.2 Dampers 1$^{st}$ generation (30 MHz)

After installation and commissioning of the dampers, the first time an instability occurred was on February 14$^{th}$, 2006, when the number of antiprotons stored in the Recycler was $> 400 \times 10^{10}$. It happened during extraction to the Tevatron, after the 2$^{nd}$ transfer. As a routine procedure, the beam was mined [5] into nine parcels with high momentum tail particles captured in a hot momentum bucket. Figure 6 illustrates this instability. It shows the beam loss, the vertical emittance jump and the vertical damper kicker output which responds to the instability. This picture is typical. More details will be shown later on a more recent instability event when more diagnostics were available. Another typical characteristic was that not all bunches lost beam.

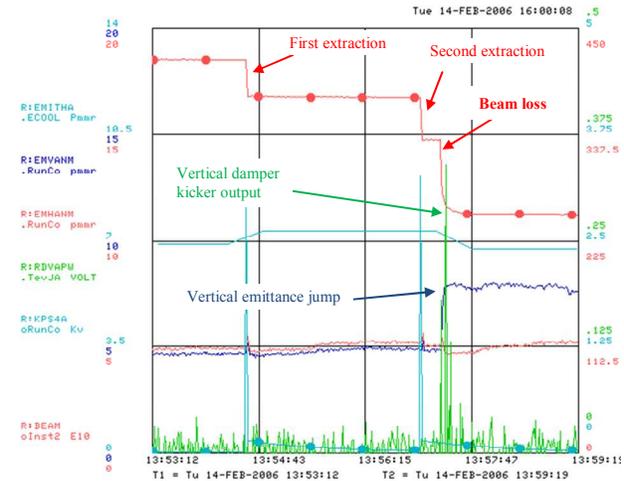

Figure 6: February 14$^{th}$, 2006 instability.

Following this instability, more efforts were undertaken to understand in more details this phenomenon and the potential limitations of the damper system. Below is a table that summarizes dedicated studies carried out in 2006-07, all in a single bunch configuration.

One peculiarity of these studies is that it was often difficult to induce an instability with the dampers on. It is possible that the electron beam performance was unequal and high enough phase densities were not achieved for the studies where the instability would not develop. On the other hand, other measurements, data taken during instabilities that occurred during normal operation and the study of 12/24/07 (for example) showed that the 1$^{st}$ generation dampers were not sufficient to handle larger stacks and that more bandwidth was needed.

In addition, an important operational limitation was found to be the saturation of the dampers' pickup preamplifiers output signal. It was observed during beam preparation for extraction, when the linear beam density increases by more than a factor of 2. Saturation was effectively turning off the dampers and the developing instability and accompanying beam loss yielded "clipping" all bunches down to the same peak density.

Table 2: List of instability studies carried out with the 1st generation dampers.

| Date | $N_p$ [$\times 10^{10}$] | Final pulse gap [μs] | Comments |
|---|---|---|---|
| 9/7/05 | 110 | 1.7 (constant) | Code bug: anti-damping turned on unintentionally |
| 2/21/06 | 58 | 0 | No instability; Dampers saturated |
| 2/22/06 | 56 | 0 | Instability during final squeeze; Dampers saturated at 0.2 μs bunch length |
| 11/7/07 | 48 | 2.5 | Small beam loss; $D_{th, 95}$ ~2.7 |
| 12/4/07 | 99 | 1.7 (constant) | Dampers gain reduced in steps; Instability only when dampers turned off |
| 12/24/07 | 342 | 8.7 (constant) | $D_{th, 95}$ ~2.6 |

## 3.3 Dampers 2nd generation (70 MHz)

The choice of an upgrade of the dampers to a bandwidth of 70 MHz was mainly dictated by the design limitations of the pickups and kickers so that all improvements could only come from upgrading the electronics [9, 10]. In addition, saturation of the dampers' pickup pre-amplifiers output signal was effectively eliminated although drifts of the trajectory within the pickups are monitored and corrected to ensure that these signals remain minimal.

Table 3: List of instability studies carried out with the 2nd generation dampers.

| Date | $N_p$ [$\times 10^{10}$] | Final pulse gap [μs] | Comments |
|---|---|---|---|
| 12/26/07 | 293 | 4.9 | $D_{th, 95}$ ~4.4 |
| 4/29/08 | 82 | 1.8 | No instability; $D_{95}$ ~4.3 |
| 1/14/09 | 53 | 0.2 | Mining-like conditions; $D_{th, 95}$ > 4 |
| 4/13/10 | 154-199 | N/A | Various RF |
| 12/27/10 | 300-415 | N/A | Various RF |

After commissioning (November 2007 – May 2008) the new dampers were declared fully operational. Table 3 summarizes the studies that have been carried out since then. The first 2 studies in the table were intended to directly compare the 2nd generation dampers with the 1st generation, in particular the studies performed on 12/24/07 (Table 2, 1st generation dampers) and 12/26/07 (Table 3, 2nd generation dampers). These measurements indicate that the dampers upgrade might have resulted in a ~70% increase of the instability threshold limit, which is in quite good agreement with the model predictions (*e.g.*: see Table 1 line 3 and Table 3 line 1; in both cases we had the pulse gap sufficiently larger than 4× the nominal pulse width of 0.9μs.).

The other studies, on which this paper focuses, are investigations related to instabilities that occurred during operational conditions.

## 4. RECENT INSTABILITIES DURING NORMAL OPERATION

### 4.1 History of instability occurrences

After the damper system upgrade mentioned in the previous section and full completion of its commissioning, we recorded 6 instabilities during regular operation over a period of about 2 years, while continuing to adjust cooling parameters and modifying procedures. Table 4 below summarizes the history of these instabilities along with relevant parameters and comments. Note that all the instabilities occurred during the extraction process and only in one specific RF configuration, the "mined" bunch (see the next section).

In Table 4, instabilities that were induced as part of a dedicated study (and not during an actual shot to the Tevatron) or for which a hardware failure was identified are not included. The changes made to the procedure were either a direct consequence of the conditions in which an instability developed (*e.g.:* adjustments to the electron beam position to reduce cooling) or attempts to improve stability (*e.g.:* removal of the high momentum bucket).

The relevance of indicating the final cooling time before an extraction to the Tevatron in Table 4 results from the observation that an instability is more likely to occur when the antiproton beam has been cooled for a long period of time without any further injections from the Accumulator (typically, the final cooling time is of the order of 1 hour). When the antiprotons remain in a single bunch configuration for several hours undergoing stochastic and electron cooling without being disturbed by the RF manipulations that take place during injections, most tails particles ought to be either brought into the core of the distribution or lost to the vacuum chamber. As a result, Landau damping is greatly suppressed and conditions for an instability to develop are enhanced.

Table 4: List of all the instabilities observed during operation since completion of the dampers upgrade.

| Date | $N_p$ [×$10^{10}$] | After bunch # | REC status | Final cooling time [hrs] | Procedure change as a result |
|---|---|---|---|---|---|
| 11/19/08 | 360 | 4,5 | | 4 | None |
| 12/7/08 | 400 | 5,6 | After electron beam tuning | 8 | Electron beam offset changed from 0 to 0.5 mm |
| 05/24/09 | 370 | 7 | | 5 | None |
| 09/27/09 | 420 | 5 | After CS alignment | 24 | Electron beam offset at extraction changed from 0.5 to 0.8 mm |
| 01/07/10 | 410 | 6 | Cooling with 300 mA 'study' during extraction | 3 | Adjusted electron beam offset for 0.3 A |
| 02/22/10 | 390 | 5 | Cooling with 300 mA 'study' during extraction | 1.3 | Removed bucket with high momentum particles |

A direct consequence was the choice made to inject more antiprotons from the Accumulator shortly before beginning the extraction procedure (only one transfer). The reasoning is that by doing so, the beam distribution is stirred enough so that tail particles are repopulated and help stability through Landau damping.

The idea of removing the RF waveform that isolates high momentum particles (so-called 'hot bucket') from the rest of the bunch has the same purpose. The 'hot bucket' was created [5] as a way to avoid unnecessary losses during extraction when only stochastic cooling was available and the amount of high-momentum particles was large. With electron cooling, emittances are now significantly lower and, typically, ~97% of the antiprotons are extracted. As a result, although removing the 'hot bucket' increases the tail population hence improves the antiprotons beam stability, we find that it does not affect the efficiency of the transfers to the Main Injector and further acceleration. Although no dedicated study was carried out, no instability has been observed during operation since the removal of the 'hot bucket'.

The last interesting information shown in Table 4 is the fact that no instability has ever been observed before any of the 'mined-bunches' was extracted. In fact, in all cases, at least 3 of the 9 'mined-bunches' had been extracted before an instability was seen. The most straightforward reason is that the last mined bunches to be extracted are cooled longer than the very first ones. Another possible factor is how DC particles contribute to Landau damping. Right before the extraction, the region where DC particles have a low momentum offset corresponds only to a small portion of the ring. Therefore these DC particles spend most of their time inside the buckets keeping them stable. In contrast, when only one bunch is left (*i.e.* close to the end of the Tevatron shot) nearly the same number of DC particles is spread over a larger longitudinal phase space resulting in a lower density in the tail particles. Consequently, the situation may not be favorable for stability, and the DC particles are effectively excluded from Landau damping.

### 4.2 Example

A typical instability is characterized by three phenomena: a large and sharp increase of the damper kickers' amplitudes (in particular, the vertical damper kicker); a fast increase of the emittances (mostly vertical) as measured by the Schottky detectors; and a relatively slow beam loss. Figures 7a and 7b shows these features for the instability that occurred on 05/24/09.

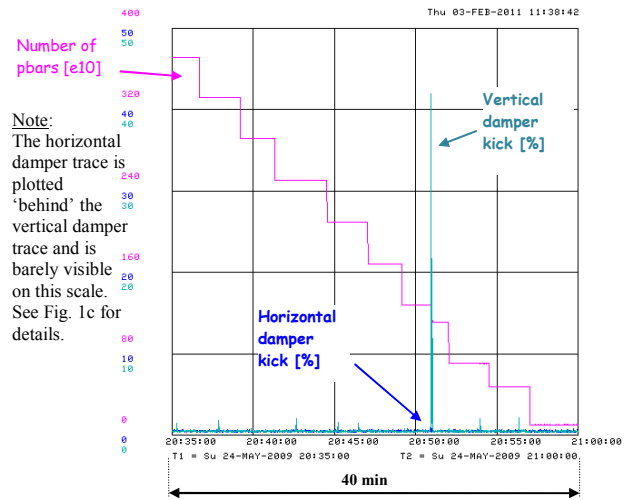

Note: The horizontal damper trace is plotted 'behind' the vertical damper trace and is barely visible on this scale. See Fig. 1c for details.

Figure 7a: Extraction sequence (05/24/09). In this case, the instability occurred after the 6[th] of the 9 mined bunches had been injected into the Tevatron.

The instability lasts for 5-15s and accordingly, the beam loss is slow, while without dampers (or with malfunctioning dampers) most of the beam loss and the emittance blow up happen in < 0.1s. It corroborates with the fact that the instability growth rate $\mathrm{Im}\,\Delta\nu_c$ for the lowest betatron sideband is ~ 40 times higher, than at 70 MHz. Not shown on Figure 7 are the emittances measured by the flying wires, which are almost unaffected by the instability, indicating that this is mostly the tail particles that suffer from the instability and are being lost to the aperture, and this is also consistent with the general picture of the Landau damping.

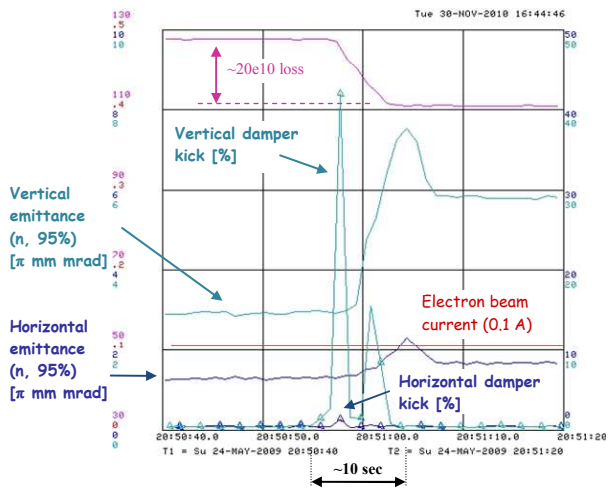

Figure 7b: Close up from Figure 7a at the time of the instability (data recorded at 1Hz).

When an instability develops, oscilloscope traces of the dampers pickup electrodes are automatically recorded for 32 ms (limited by the oscilloscope memory capacity). The sum and difference signals (proportional respectively to the current density and the beam position), from the vertical damper pickup electrodes are shown on Figure 8 for the event from 05/24/09. The two plots show how the oscillations amplitude grows during the instability. Other features are that this is always the trailing edge of the bunch that goes unstable and that the instability does not propagate as it develops.

## 5. DEDICATED INSTABILITY STUDIES

### 5.1 RF profiles during extraction

The extraction process requires complicated RF gymnastics [5] that primarily uses 3 distinct RF configurations:
- single barrier bucket bunch before the 'mining' process (called 'cold bucket' configuration);
- 'mined buckets' configuration, which consists of up to 9 short bunches within rectangular RF barriers;
- '2.5 MHz buckets', which is composed of four 2.5 MHz bunches ready to be extracted.

Initially, the beam is kept in a long cold bucket. First, the bunch is divided into 9 nearly identical pieces with narrow rectangular barriers (called for historical reasons "mined bunches"). Then antiprotons are moved, one mined bunch at a time, into the extraction region. Once there, the mined bunch is adiabatically transformed into four 2.5 MHz smaller bunches, which are then extracted into the matching MI RF waveform.

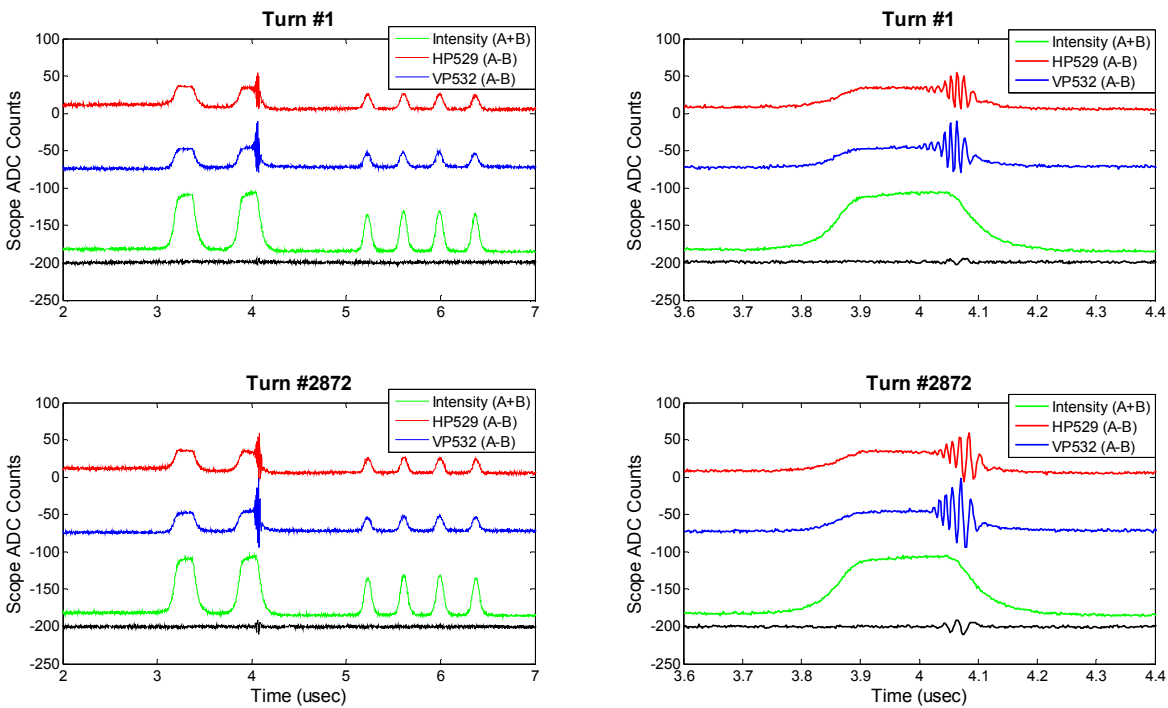

Figure 8: Oscilloscope traces from the damper pickups of the 05/24/09 instability. Left: 5 μs (out of 11.1 μs for the whole circumference); Right: zoomed on the bunch that went unstable (Bunch #8). The green trace is the sum signal and is proportional to the linear density distribution. The red and blue traces are the differential (not normalized) signals and reflect the beam transverse position in the horizontal and vertical directions, respectively. The black curve is the dampers kick. Top plots: beginning of the instability; Bottom plots: end of the recording period (32 ms). Other bunches did not show any oscillations in the recorded set.

Two of the RF configurations are presented in Figure 9, along with the corresponding effective potentials calculated from the RF fan back and the beam longitudinal profile (*i.e.* linear density) from the resistive wall monitor (RWM).

Missing from Figure 9 is the 'cold bucket' waveform for which the potential well is depicted in Figure 4. It is our standard configuration during accumulation and consists simply of the beam contained within two rectangular RF pulses.

## *5.2 January 14, 2009, study*

The goal of this study was to mimic a single mined bunch and find the instability limit experimentally. For this purpose, $53 \times 10^{10}$ antiprotons (close to the typical intensity of a single mined bunch) were captured in a single barrier bucket, 0.68 μs wide (*i.e.* thirty six 53 MHz-RF buckets: two thirteen 53 MHz-RF bucket wide RF pulses with ten 53 MHz-RF bucket space, the operational width of a mined bunch). Because the importance of the finite depth of the potential well was not recognized at that time, the barriers had the 'standard' height and width as those of the cold bucket (0.9 μs, 1.8 kV RF amplitude, which corresponds to a barrier height of 17 MeV/c). The antiprotons were cooled down with the electron beam until an instability developed. After ~15 minutes of cooling at the maximum strength, $D_{95}$ reached 4.7, the beam went unstable, and $\sim 35 \times 10^{10}$ antiprotons were lost. Figure 10 shows several parameters at the time of the instability. The evolution of the beam parameters that are common to Figure 7b and Figure 10 looks very much alike. In particular, like for the operational event shown in the preceding chapter, the instability (and beam loss) lasts a relatively long time, ~15 s in this case.

## *5.3 April 13, 2010, study*

### 5.3.1 Scheme of the study

For this study, the instability threshold limit for each of the 3 RF waveforms listed in Sec 5.1 was investigated. For the cases involving the 2.5 MHz structure, we used a novel idea where only a portion of the beam is subject to go unstable. In order to reach conditions close to those encountered during normal operation, the initial bunch is sliced into two parts so that the number of antiprotons to which the relevant waveform will be applied is equivalent to the number of protons we would get during a normal extraction of $400 \times 10^{10}$ or so particles. The beam portion that is supposed to stay intact is kept in a cold bucket which occupies as much of the ring as possible. Two factors allow separating the thresholds of these two portions of the beam. First, the efficiency of electron cooling drops when the antiproton momentum spread is below ~4 MeV/c. Because antiprotons in the cold bucket have a lower momentum spread, hence are cooled less effectively, their phase density stays lower. Second, the long cold bucket configuration is beneficial for keeping all resonant particles within the bucket.

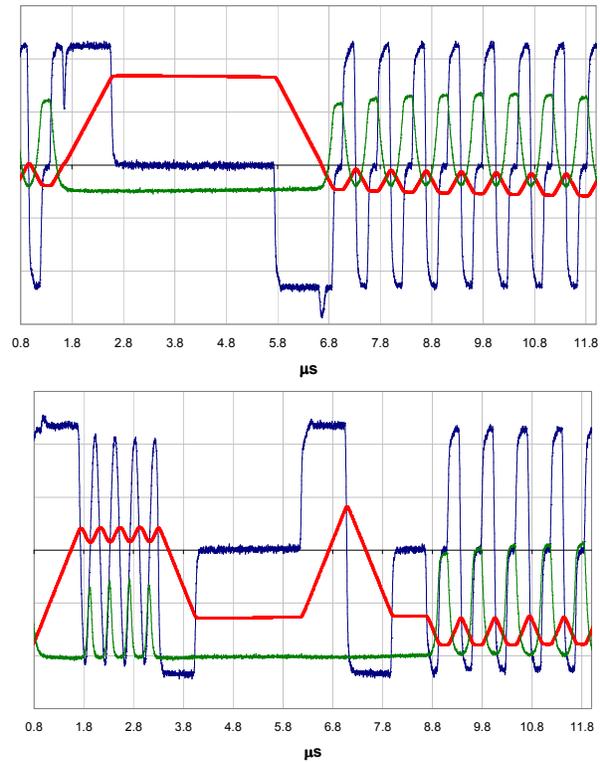

Figure 9: 'Mined buckets' configuration (top) and '2.5 MHz buckets' configuration (bottom). Blue trace: RF fan back; red trace: effective potential; green trace: longitudinal beam density measured with the Resistive Wall Monitor (RWM).

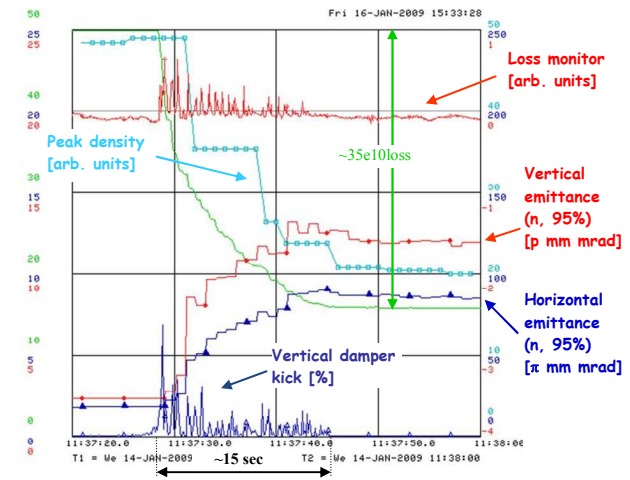

Figure 10: Instability from 01/14/09.

Finally, the dampers pickup electrodes oscilloscope traces recorded during the study are basically indistinguishable from those displayed on Figures 7a and 7b.

### 5.3.2 Diagnostics

One of the complications in the study was the difficulty of using standard emittance measurement tools. In all previous instances, the value of $D_{95\%}$ was calculated using the average transverse emittance, (H+V)/2 and the longitudinal emittance. The longitudinal emittances are based on un-gated signals from 1.76 GHz Schottky pickups and the beam in a standard rectangular barrier bucket with RF pulses width of 0.9 µs and height of 1.8 kV. In the case of a complicated RF waveform, this method cannot provide a correct answer, and the measurement was done off-line in an alternate way [13].

The transverse emittance was measured with the horizontal flying wire with the beam profile fitted to a Gaussian function for the calculation of its width (at the time of the study, the vertical flying wire scanner was out of order). The profiles were recorded with the signal gated over the portion of the beam of interest. Note that in the case of a long cold bucket the flying wire emittance was always lower than the Schottky's by at least a factor of 1.2 (likely a calibration issue), and the ratio increased by up to a factor of 1.8 when the beam was deeply cooled by the electron beam, indicating the formation of long non-Gaussian tails. Therefore, the numbers for $D_{95\%}$ quoted in this section are consistent within this section of the paper but may be larger up to a factor of 1.8 than obtained with the standard tools, used in other sections.

The longitudinal emittance was calculated with the tomography procedure [13] applied to the longitudinal density profile, which was acquired with a RWM together with the corresponding RF voltage waveform. The tomography analysis consists in taking the waveforms of the RF voltage and of the beam longitudinal current density (from the RWM) to construct a plot of the percentage of the beam outside a given phase space area as a function of this phase space area. An example of the output plots obtained through this method is shown on Figure 11. From this plot, one can read the longitudinal emittance by finding the phase space area that corresponds to a given percentage of the beam. For instance, the 95% longitudinal emittance from Figure 11 is 1.2 eV s.

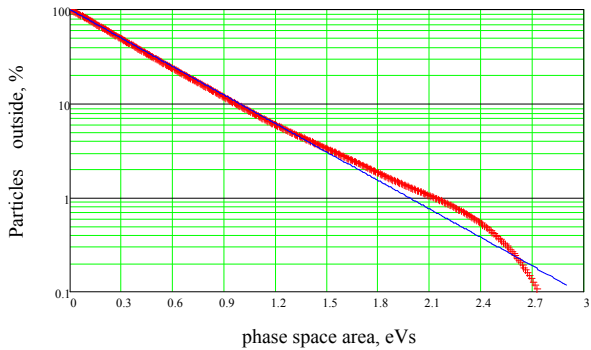

Figure 11: Phase space integral for the 2nd bunch of the 2.5 MHz buckets case.

During the tomography analysis, it was realized that the RF and RWM waveforms were not exactly synchronized which would result in some systematic error. However, this error can be corrected by taking into account the fact that the linear density is a unique function of the potential, hence, the left side and the right side of a bunch profile must give the same dependence as a function of the potential. A 5 ns timing adjustment took care of this issue. The tomography approach gave the same results as the calculation obtained from the Schottky signal for not-too-deeply cooled bunches contained between rectangular barriers.

As for all instability events, an oscilloscope was connected to the dampers pickup electrodes. It was triggered by a high transverse signal if it occurred above 70 MHz and recorded 32 ms of data.

### 5.3.3 Example of an instability event during the study

Several features were common for all instability cases:
- duration of the beam loss (10-15 sec);
- response primarily from the vertical damper;
- emittance growth primarily in the vertical plane;
- length of the beam affected by the transverse motion (100 – 200 ns) over the 32 ms of recorded data on the oscilloscope.

Plots below are shown for the "2.5 MHz buckets with anti-barriers" configuration but are illustrative of all cases. Figure 12 is equivalent to Figure 7a, and shows relevant parameters during the instability.

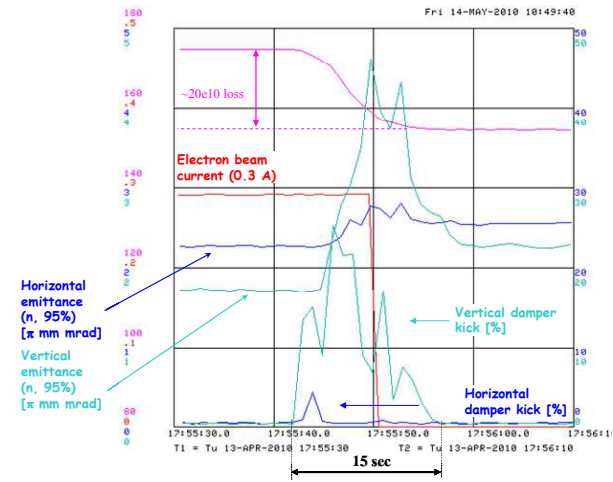

Figure 12: Example of an instability event for the "2.5 MHz buckets with anti-barriers" configuration.

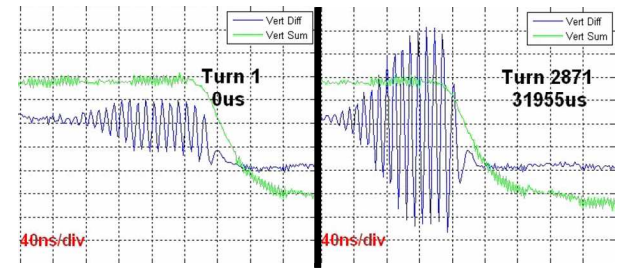

Figure 13: Oscilloscope traces of an instability for a single bunch 0.94 µs long. The vertical scale is arbitrary. The green trace is the line density distribution. The blue trace shows to the beam oscillations during the instability. Left plot is at the beginning of the instability. Right plot is at the end of the recording period.

Similarly, Figure 13 shows the sum and difference signals from the vertical damper pickup electrodes for a typical instability event encountered during the study.

It is clear by comparing Figure 7a with Figure 12 and Figure 8 with Figure 13 that the characteristics of the instability are almost exactly the same, thus validating the study procedure.

As mentioned previously, the instability does not propagates through the bunch in the time recorded by the oscilloscope as it develops and, in the same manner, when several bunches are present, the instability affects only one bunch, leaving the other bunches unaffected. This is illustrated in Figure 14. There, only the 2$^{nd}$ bunch goes unstable.

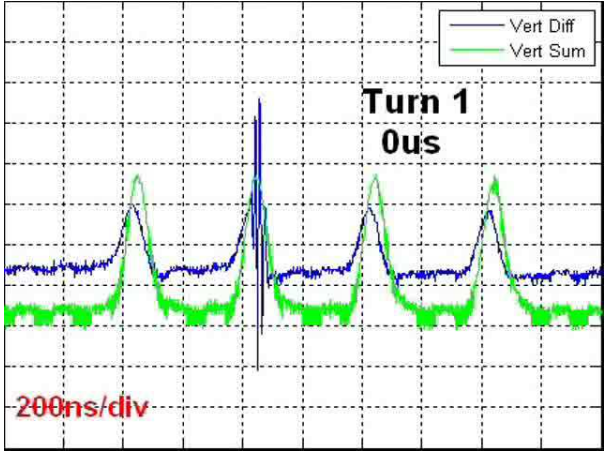

Figure 14: Oscilloscope traces of an instability for the "2.5 MHz buckets" configuration. Traces are the same as for Figure 13.

5.3.4 RF configurations

As mentioned previously, one of the goals of the study was to identify which RF configurations from the extraction RF manipulations are the most and least robust against the instability. Thus, the 3 cases independently investigated were: the 'cold bucket' configuration, the 'mined buckets' configuration and the '2.5 MHz buckets' configuration. In addition, one alternative case was tried but will be discussed separately. Note that the "Case" numbers below are simply a representation of the order in which the study was conducted and are mentioned for internal purposes only.

A. *'Cold bucket' configuration (Case #4)*

During extraction, the 'cold bucket' contains a single bunch 6.11 μs long between two RF barriers. For the study, because the number of antiprotons available was limited ($154 \times 10^{10}$), the barrier pulse gap had to be reduced to 0.94 μs (or fifty 53 MHz buckets) in order to get the same peak current density as for a normal extraction to the Tevatron. The barrier width is 0.9 μs (or forty eight 53 MHz buckets) and the RF height is maximum (1.8 kV, measured on 09/14/2009). Figure 15 shows the RF voltage, the corresponding potential and beam profile from the RWM. The slope on the bunch profile comes from imperfections in the RF system which cannot deliver a perfectly flat potential well and from beam loading.

Just before the instability, the 'on-line' phase density reached 4.5 and the total beam loss was $42 \times 10^{10}$.

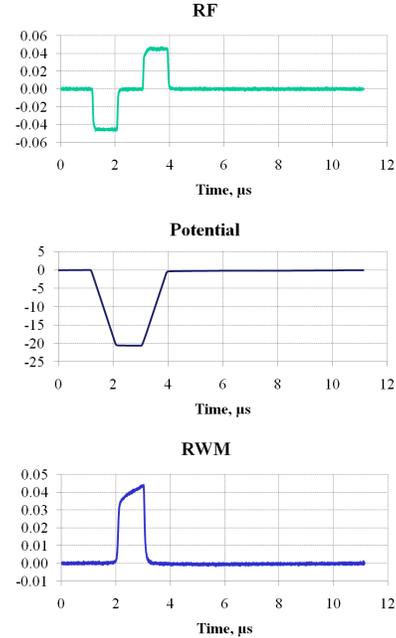

Figure 15: RF fan back (top), equivalent potential (middle) and beam linear density for the RWM (bottom) for the 'cold bucket' configuration. Vertical scales for the RF fan back and the RWM are Volts (raw units from the oscilloscope).

B. *Mined bucket configuration (Case #1)*

The first step of the extraction process is a manipulation called mining. At the end of this manipulation, the single bunch has been converted into 9 short bunches (mined bunches) which overall occupy the same space in the ring. During normal operation, the barriers width of a mined bucket is 0.68 μs and the RF voltage is ~1.8 kV with a bucket height ~ 9.1 MeV. In this study we chose exactly the same bucket height as the one used for the operational mined bucket. This was accomplished by reducing the barrier pulse height with respect to its value set for the configuration shown in Figure 15. As a consequence of this, the length of the well's bottom (fifty vs. ten 53 MHz-RF buckets) was five times longer than that used during normal mining. In this case, the 'on-line' phase density just before the instability was 6.9 and the total beam loss $21 \times 10^{10}$ from $199 \times 10^{10}$ before the instability.

C. *'2.5 MHz buckets' configuration with anti-barriers (Case #3)*

After the beam has been mined, one by one, each mined bucket is brought into the extraction region of the ring and morphed into four 2.5 MHz bunches, which potential is elevated with respect to the rest of the beam by so-called 'anti-barriers' to avoid transferring the DC particles. Figure 16 shows the RF voltage, the corresponding potential and beam profile from the RWM. Note that in this case, the portion of the beam which is not contained in the 2.5 MHz buckets occupies the remainder available space in the ring.

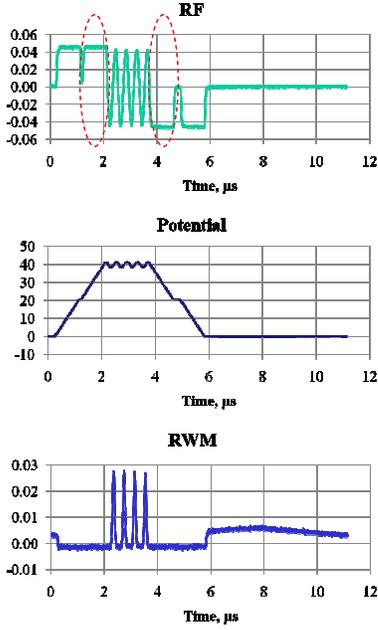

Figure 16: RF fan back (top), equivalent potential (middle) and beam linear density for the RWM (bottom) for the '2.5 MHz buckets' with anti-buckets configuration. The dotted red ellipses on the top plot indicate the so-called anti-buckets.

For this case, the total number of antiprotons is $175\times10^{10}$ but only $42.2\times10^{10}$ is contained in the 2.5 MHz RF structure (calculated from the RWM waveform). As explained previously, because the online calculation of the longitudinal emittance assumes a single bunch structure, which is obviously not correct in this case, the online phase density at the onset of the instability was not available. The total beam loss was $20\times10^{10}$ but only 3 out of the 4 bunches were affected. On the RWM it resulted in 3 of the bunches being clipped (*i.e.* smaller peak current density than the 4th bunch).

D. *'2.5 MHz buckets with barriers' configuration (Case #2)*

This RF configuration was an attempt to assess the benefit/drawback of the elevated potential during extraction. The only difference with the previous case is that the anti-buckets were replaced by normal barriers as illustrated in the top plot of Figure 17 (to be compared with the top plot of Figure 16). The result of flipping the RF barriers that surround the 2.5 MHz structure is that it places the 2.5 MHz bunches at the bottom of the potential well instead of the higher potential they were at in the previous case.

Both the total number of antiprotons ($176\times10^{10}$) and the number of antiprotons contained in the 2.5 MHz structure ($54.3\times10^{10}$) are similar to the previous case. Although signs of an instability were observed by a large vertical damper kick and emittance growth, it seems to have been of a different nature than all other cases. The oscilloscope was not triggered, indicating that the frequency was lower than 70 MHz. Also, less than $1\times10^{10}$ was lost. Because the reason for the instability is unclear, we cannot compare it to Cases A, B and C. Consequently, the analysis was not carried out for this case.

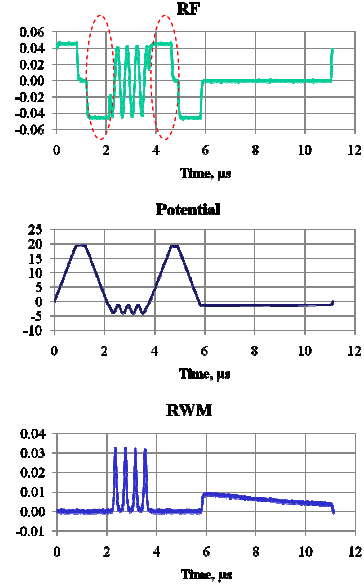

Figure 17: RF fan back (top), equivalent potential (middle) and beam linear density from the RWM (bottom) for the '2.5 MHz buckets' with cold bucket configuration. The dotted red ellipses on the top plot indicate the barriers that form the 'cold bucket'. These barriers are flipped with respect to the anti-buckets configuration.

5.3.5 Results of the study

The recorded data were analyzed as outlined in the first paragraph of this section. From the RF and RWM waveforms, the longitudinal emittance just before an instability was computed using a tomography procedure. Then, using the transverse (vertical) emittance measured at the same time with a flying wire detector, and the number of antiprotons in the bunch considered, the phase density $D_{th,\,95}$ was calculated and chosen as the threshold limit for a given RF configuration.

The results are summarized in Table 5, in which Case D (paper nomenclature) has been omitted due to its 'non-standard' characteristics. The longitudinal emittance in Table 5 and the transverse emittance from the flying wire measurement (not reported here) are both 95% normalized emittances.

In Table 5, for Case C, the threshold phase density $D_{th,\,95}$ is actually the average phase densities of each individual 2.5 MHz bunches which were computed independently. For bunches 1 to 4 (left to right on the RWM waveform from Figure 17 for instance), the phase density (95%) is respectively 7.1, 7.4, 7.4 and 6.9. Coasting Gaussian beam model, Eq. (6), gives $D_{th,\,95} = 5.5$.

Table 5: Summary of results. Case numbers use the paper nomenclature.

| Case # | Total # pbars [$\times 10^{10}$] | Number of pbars from RWM [$\times 10^{10}$] | Longitudinal emittance [eV s] | Phase density $D_{th,\,95}$ |
|---|---|---|---|---|
| B | 199 | | 19.5 | 3.8 |
| C | 175 | 42.2 | 1.2 | 7.2 |
| A | 154 | | 9 | 10.9 |

The phase densities listed in Table 5 show that the threshold limit for Case B is lower than for Case C, which is lower than for Case A. In other words, the 'cold bucket' configuration is the most robust against the instability, while the 'mined bucket' configuration is the most prone to become unstable. This is consistent with the historical observations during normal operation where this is always one of the nine 'mined' bunches that went unstable.

## 5.4 December 27, 2010, study

In essence, the intent of this study was very similar to that described in Sec 5.3. The first set of measurements was dedicated to the investigation of the impact that the depth of the potential well has on the beam stability. The second measurement was simply a repeat of Case D (Sec. 5.3.4 D), for which results were inconclusive. In addition, while for the April 13, 2010, study, the number of antiprotons available was limited and lead to non-standard RF manipulations, the number of antiprotons for this study was large (close to nominal), which simplified the beam preparation.

Note that, $D_{th,\,95}$ reported in this section is the 'online' phase density, which is calculated with the Schottky measurements for the emittances and not with the tomography procedure and the flying wire as it was for Table 5. Using the tomography analysis and flying wire data always results in phase density threshold higher than the 'online' calculation by a factor up to 1.6 while showing the same trends.

### 5.4.2 Observations and results

*1  Two different potential depths*

For this study, the 'cold' bucket configuration was chosen (as in Sec. 5.3.4 A). The barrier pulse gap was three hundred and eight 53 MHz buckets, the same length as for operation during normal accumulation. The barriers width was also standard at forty eight 53 MHz buckets. In the first measurement, the full RF voltage (~1.8 kV) was applied and the number of antiprotons was $415\times10^{10}$. The beam went unstable at $D_{95}\sim 5.5$. For comparison, in the case described in Sec. 5.3.4 A the 'online' phase density threshold was 6.8. The two measurements are thus consistent.

However, this instability had unique features. First and foremost, the beam oscillations did not occur at the tail of the bunch but near the middle. Figure 18 shows the dampers pickups recorded during the event. Too high a value of the diagnostics' low frequency cut-off does not allow resolving the details of the longitudinal density distribution, but from other measurements it is known that RF distortions result in a minimum of the RF potential in the middle of the beam. When the beam is cold, the potential minimum reveals itself in a slight peak of the linear density. In this specific case of a long bunch, it might create a preferential condition for the instability.

Several peculiar features can be seen on the data displayed in Figure 18. First, the oscillations are extremely local in space. Second, their amplitude increases in space (along the horizontal axis of Fig.18) much faster than it decays. Third, a numerical comparison of intensities in the first and the last turn shows a ~10% dip in the location of the (presumably) highest amplitude. Therefore, the loss seems to stay very much localized even in a case of a continuous bunch, similar to a multi-bunch case illustrated in Fig.14. These features seem to be in agreement with the almost local character of the stability threshold, Eqs (1) and (6), where a single global value of the coherent tune shift enters only logarithmically. The effective density is not quite the same along the bunch, since the potential well is not quite flat. When the beam is cooled, its highest effective local density slowly crosses the threshold, and gets unstable, while the most part of the bunch is still below the threshold. Thus, it seems to be reasonable to expect the local character of the instability. One more factor which works on the instability localization is image currents and charges. Indeed, coherent betatron frequencies of different parts of the bunch are not the same due to the detuning quadrupolar wake [2] and some variations of the bunch linear density along its length. Small uncontrolled variations of the almost constant linear density may result in significant changes of the spatial pattern of the unstable mode. A significant spatial asymmetry of this pattern can be expected due to causality of the wake function.

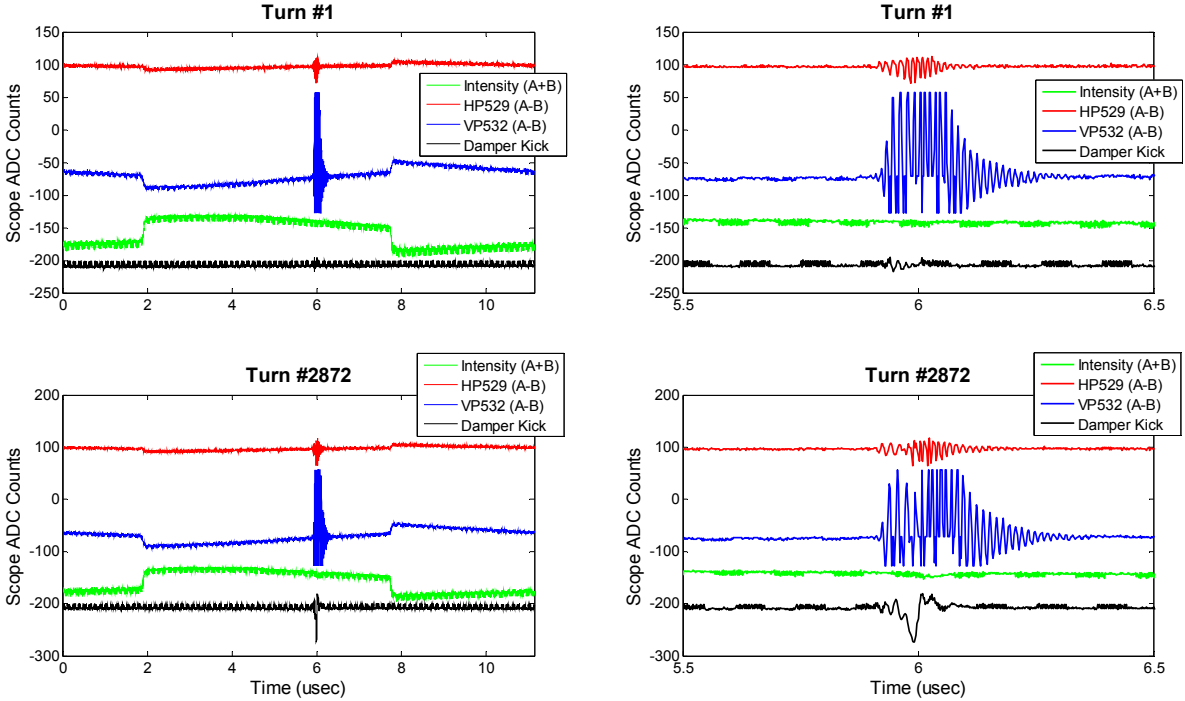

Figure 18: Dampers pickups oscilloscope traces captured at the time of the instability (1$^{st}$ turn – top; Last turn (after 32 ms) – Bottom). Plots on the right are a close up view near the location of the instability within the bunch. The legend is the same as for Figure 8. Note the fast rise and slow decay of the oscillations. Also barely visible is a ~10% dip of the beam intensity where the instability occurs. The periodic noise on all traces and the apparent saturation of the vertical pickup (VP532) are oscilloscope artifacts.

For the second part of the measurements, the RF voltage was reduced to 20% of its nominal (and previous) value lowering the potential depth by 2.2 times. The number of antiprotons was $328 \times 10^{10}$. Three distinct instability-like events occurred with several minutes intervals. The first two had very small consequences: beam loss of $3 \times 10^{10}$ and $1 \times 10^{10}$, respectively; almost no emittance growth (as opposed to all other measurements). Because of these characteristics, cooling was not halted and, eventually, a more "standard" instability developed *i.e.* fairly large emittance growth and $10 \times 10^{10}$ antiprotons lost. For this case, the instability threshold was 5.0 -5.3, similar to what has been observed for the full-height barriers in the previous experiment of the same day (Sec. 5.4.2.1). A possible explanation of discrepancy with results of the April 13, 2010 study can be related to the difference in the barrier pulse gap. Here the bunching factor *B* (the ratio of the revolution period to barrier-to-barrier time) was $B = 1.9$, while in the April 13, 2010 study B = 12 When a bunch is squeezed with unchanged depth of the potential well, the Landau resonant particles may leave the bucket, leading to a strong reduction of the Landau damping, and thus, lowering the threshold density. However, when the gap is sufficiently small, the resonant particles return to the bucket very soon, and Landau damping almost does not suffer. Note again that the tomography analysis was not carried out for this set of measurements as opposed to the values reported in Tables 5 and 6.

*2  2.5 MHz buckets with barriers' configuration (similar to Sec. 5.3.4.D)*

This measurement was a repeat of the one made on April 13, 2010, and which was inconclusive as per the characteristics of the stability then observed. Except for the number of antiprotons that remain in the 'cold bucket', all other parameters were basically equal (Note: the number of antiprotons captured in the 2.5 MHz was $\sim 50 \times 10^{10}$, but could not be measured precisely). The instability did not develop even with electron cooling in its strongest possible configuration. While it can be considered as an indication that this configuration is the least prone to instability, no specific number for the phase density could be measured.

### 5.5 Data from a normal extraction

While this is not done routinely during normal operation, the same analysis can be applied and the corresponding calculation of the phase density for various portions of the beam during extraction (*i.e.* different RF structures) can be calculated. This has been done during a typical extraction to the Tevatron and the phase densities that were obtained are reported in Table 6 along with the corresponding limits obtained from the study.

Table 6: Instability threshold limits obtained during the study and phase densities recorded during a normal extraction (hence not accompanied with an instability) for 2 RF configurations. The operational case corresponds to TeV shot #7770 (4/27/2010).

|  | Study (Table 5) – Instability threshold | Normal extraction – No instability |
|---|---|---|
| 'Mined bucket' configuration | 3.8 | 4.5 |
| '2.5 MHz buckets' configuration | 7.2 | 3.2 |

*Note: For the operational case, the data for mined bunch #9 and the average of all four 2.5 MHz bunches of mined bunch #8 are presented*

Within the level of accuracy of the measurements compiled in Table 6 (dominated by the difficulty of controlling the distribution tails), the phase densities for the 'mined bucket' configuration indicate that the antiproton beam is likely close to going unstable during the extraction of the last bunches for a typical extraction to the Tevatron. On the other hand, with a phase density for the instability threshold more than twice the phase density calculated during normal operation, the '2.5 MHz buckets configuration' does not pose any concern from an instability point of view. However, it should be noted that the RF waveforms for the study and during a real extraction sequence are not exactly the same (Figure 16 vs Figure 9) and given the nature of the instability mechanism, this difference might be sufficient to explain the somewhat higher value of the phase density obtained for an operational case with respect to the threshold obtained during the study.

### 5.6 Note about chromaticity

Theoretical estimations in Section 2 are given for a chromaticity $\xi = -6$, because it was the nominal value in both planes during the initial stages of the Recycler operation. In the time when most of instability studies were performed, chromaticity measurements were not done, while adjustments of the working point, trajectory drifts etc. over the years were likely changing the "natural" chromaticity. However, it has been observed that the natural chromaticity of the Recycler had a time dependent drift. The recent measurement of the chromaticities (January 2011) gave $\xi_H = -2$ and $\xi_V = -4$, significantly different from the assumed value. This uncertainty with the chromaticity value during the experiments might have added to the scatter in the instability threshold results. Nevertheless, note that, in accordance with Eq. (1), the higher the bandwidth of the dampers, the less impact the natural chromaticity has on the instability threshold and, in particular, a factor of 2 does not imply such a difference for the stability threshold.

## 6. SUMMARY AND DISCUSSION

### 6.1 Summary of the experimental results

The instability thresholds recorded with Schottky emittances are summarized in Table 7.

Table 7: Summary table of the experimentally observed instability thresholds. The phase density is calculated with Schottky emittances.

|  | $D_{th,95}$ |
|---|---|
| No dampers ($n = 0$) | 0.5 – 0.8 |
| 30 MHz dampers ($n \sim 330$) | 2.6 - 3.1 |
| 70 MHz dampers ($n \sim 780$) | 3.4 – 6.2 |

The data have a large scatter and depart from the numbers calculated in Table 1. The uncertainty with the actual chromaticity, deviation of tails from Gaussian, and the effect of the depth of the longitudinal potential well are likely the main contributors to the scatter and discrepancies. On the other hand, the model correctly describes the trend of introducing dampers with different bandwidths, and the observed thresholds are within a factor of two of the predicted numbers. Also, the instability was always at the upper edge of the dampers bandwidth.

### 6.2 Interpretation

The detailed studies with the final version of the dampers showed a qualitative agreement with the prediction of the lower instability threshold for bunches kept in shallow potential wells.

Several features were common for all recently observed instability cases:

1. The instabilities occur primarily in the vertical plane.
2. No dramatic changes in the oscillations characteristics were identified in the recorded 32 ms-long damper pickup signals. The high-amplitude oscillations stay localized within a 100 – 200 ns region around the highest beam linear density; the oscillations amplitude grows only by < 50%; there are no significant changes in the beam intensity (*i.e.* the beam loss occurs primarily later).
3. Likely related to point 2 was the long (many seconds) duration of the beam loss. One can speculate that the large beam loss occurs only after the resonant particles are lost, which happens in a sub-second time. In this scale, the synchrotron motion is already important and causes large deviation from the coasting beam model described in Section 2.
4. As it has been mentioned before, the threshold phase density for "2.5 MHz" RF configuration was twice higher than for the "mined bucket" configuration. Because the measurements were performed with the same dampers and the beam parameters did not differ dramatically so that it would modify significantly the logarithm in Eq.(6), we interpret the result as an effect of excluding longitudinal tail particles from Landau damping as a result of decreasing the depth of the potential well.

The features presented in points 2 and 3 (slow non-exponential growth of the oscillations and seconds-long times beam losses) should not be surprising. Indeed, a classical exponential growth of instability describes a system sufficiently above the threshold, while in all our experiments the beam was slowly reaching the threshold as it was being cooled. Strictly speaking, the instability growth rate at the exact threshold is zero. Then, in this case, it is determined by such factors as beam cooling, synchrotron motion and all sorts of diffusion for the resonant particles. That is why for that gradual approach of the threshold, the emerging instability can be orders of magnitude slower than the pure impedance-related growth.

### 6.3 Operational consequences

The brightness of the antiproton beam at the time of extraction from the Recycler is limited by the instability discussed in this paper. Several possibilities to increase the brightness have been considered.

The first one was an improvement to the existing extraction process - the original process was a separation of the particles with large longitudinal action from the rest of the bunch [5]. This procedure was developed in order to "mine" out only the "cold particles" at a time when only stochastic cooling was available in the Recycler and it allowed decreasing the longitudinal emittance of the extracted beam. With efficient electron cooling, the limitation for the extracted emittance is the resistive wall instability discussed here, and removing the longitudinal tail particles made the instability threshold likely to decrease. When this mechanism was recognized, the step separating the tail from the core during the Tevatron shot was eliminated (last line in Table 4). While no dedicated studies were performed, operationally it allowed applying stronger electron cooling, which led to the highest phase density of extracted beams achieved at the end of Run II.

Another idea comes from the fact that all instability cases during extraction occurred in the second half of the process, even though the reported emittances were already at equilibrium values. One of the possible explanations is the importance of the RF waveform outside of the bucket containing the beam approaching its instability threshold. The model developed in this paper assumes that a particle with its momentum offset larger than the potential well's depth (as depicted in Figure 4) is excluded from Landau damping. However, it is valid only if there are large sectors of the ring where the particles spend significantly more time than above the bucket. This description is applicable to the RF waveform before extraction of the last parcel. However, in the time of extraction of the first parcels, portions of the ring with a high potential with respect to the well's bottom are narrow, hence the high momentum particles can effectively participate in Landau damping. If our interpretation of the different results obtained for short and long buckets (Section 5.4.2) are correct, the instability threshold between the first and last parcel extraction can be significantly different too. Correspondingly, the proposal was to modify the RF waveforms so that the high-potential areas available for resonant particles would be minimized at all stages of the extraction process. This idea was never implemented because it was devised too close to the end of Run II.

Finally, we considered adjusting the chromaticity to increase the stability threshold in accordance with Eq. (6). However, significantly increasing the chromaticity absolute value is harmful for the life time, and the life time is of primary importance because of the long accumulation cycle (15-18 hours). Adjusting the sextupoles right before beginning the extraction manipulations, when the beam is more prone to instabilities, looked feasible and likely beneficial, but again was not implemented because of the end of Run II.

## 7. CONCLUSION

The transverse instability of the antiproton beam in the Recycler was the final limiting factor to the brightness of the extracted beams that could be achieved. Nevertheless, the transverse dampers in conjunction with electron cooling permitted to increase the beam brightness by an order of magnitude.

Qualitative features of the measured instances of the instability fit reasonably well the model developed for a coasting beam. The onset of the instability is determined by the threshold phase density, which value is in agreement with the model within the scatter of experimental data and the precision to which this theoretical threshold can be calculated. The scatter in the data is likely related to variations in the distribution of the tails particles, which determine Landau damping. In particular, lowering the potential depth of the barrier bucket effectively excludes part of the longitudinal tails from damping and may decrease the threshold density by a factor of two.

## ACKNOWLEDGMENTS

Martin Hu and Sergei Nagaitsev made the measurements at the initial stages of the Recycler operation and set up corresponding techniques. Kiyomi Seiya participated in several instability studies, preparing complicated RF forms.

We are grateful to the entire Main Injector and Support departments for helping with many aspects of the work and measurements discussed in this paper and countless productive discussions. One of the authors (LP) would like to thank V. Balbekov for fruitful discussions about the theoretical model and its limitations.